\documentstyle[12pt,aasms4]{article}

\begin{document}

\noindent
{\it Dissertation Summary}

\begin{center}

\title{\large \bf Spectrophotometric Evolution of Old Stellar Systems}

\end{center}

\author{Hyun-chul Lee}

\affil{Center for Space Astrophysics, Yonsei University, 134 Shinchon,
Seoul 120-749, Korea}

\begingroup

\parindent=1cm

\begin{center}

Electronic mail: hclee@csa.yonsei.ac.kr

Thesis work conducted at: Department of Astronomy, Yonsei University 

Ph.D. Thesis directed by: Young-Wook Lee ;  ~Ph.D. Degree awarded: 
2001(February)

{\it Received \underline{\hskip 5cm}}

\end{center}

\endgroup


Theoretical integrated spectrophotometric quantities, 
such as spectral line indices in the Lick system (H$\beta$, Mg$_{2}$, Mg $b$, 
Fe5270, Fe5335), broadband color indices (in the $UBVRIHK$ and 
Washington $CMT_{1}$ bandpasses), and surface brightness fluctuation (SBF) 
magnitudes and colors, ranging from far-UV to near-IR have been 
consistently computed for simple stellar populations from our 
evolutionary population synthesis code. The age range of $-$ 4 Gyr $\leq$ 
$\Delta$t $\leq$ + 4 Gyr is explored, where $\Delta$t = 0 Gyr is the currently 
favored mean age of inner halo Galacitc globular clusters (GCs)
(Galactocentric radius $\leq$ 8 kpc), $\sim$ 12Gyr.
The main aim of this dissertation is to investigate
the effects of post-red giant branch (post-RGB) stars on the integrated 
spectrophotometric quantities after employing detailed 
systematic variation of horizontal-branch (HB) morphology with age and 
metallicity. 

H.-C. Lee, S.-J. Yoon, \& Y.-W. Lee (2000, AJ, 120, 998) 
showed that the popularly
used age indicator H$\beta$ index is significantly affected by the 
presence of blue HB stars. As a matter of fact, because of the systematic 
HB morphology variation, it is found that H$\beta$ does not 
monotonically decrease as metallicity increases at given ages 
but shows a kind of wavy feature. This is because there is a H$\beta$ 
enhancement due to blue HB stars reaching a maximum strength 
when the distribution of HB stars is centered around 9500 K, 
the temperature at which the H$\beta$ index becomes strongest. 
Comparison of Keck observations of the globular cluster systems 
(GCSs) in the Milky Way, NGC 1399, and M87 with our new models shows 
that a systematic shift in the H$\beta$ versus metallicity plane is 
explained if the mean age of GCSs in giant elliptical galaxies is, 
on average, a couple of billion years older than the Galactic 
counterpart. If our age estimation is confirmed, this would imply that 
star formation in denser environments proceeded much
more rapidly and efficiently, so that the initial epoch of 
star formation in more massive (and denser) systems took place 
a couple of billion years earlier than in the Milky Way.

Investigation of integrated broadband colors also shows that 
similar wavy features appear among temperature-sensitive colors, such as 
$B$ $-$ $V$. Calibration of these model results, especially 
$B$ $-$ $V$, $V$ $-$ $I$, $C$ $-$ $T_{1}$, and $M$ $-$ $T_{1}$, 
using carefully selected 
Galactic GCs (i.e., $E$($B$ $-$ $V$) $<$ 0.2) 
is quite remarkable in the sense that the inner halo 
Galactic GCs do appear older than the outer halo counterparts.
It is also noted that $C$ $-$ $T_{1}$, the well-known 
highly metallicity sensitive color index, 
should be used with great caution as a color-metallicity transformation 
relation if there are age variations within a globular cluster system.
Along with H$\beta$ and some temperature-sensitive optical broadband 
colors, the use of far-UV to optical colors is proposed as 
the best probe of age for old stellar systems using their 
HB morphology variation.

SBF colors, such as $\bar{V}$ $-$ $\bar{I}$ or 
$\bar{V}$ $-$ $\bar{K}$ are found to be almost  
unaffected by the HB morphology variation because their RGB stars are 
much brighter than HB stars. However, because HB stars are fairly 
bright sources compared to RGB stars in the shorter wavelength range, 
$\bar{U}$ $-$ $\bar{B}$ and $\bar{U}$ $-$ $\bar{V}$ are severely 
influenced by blue, hot HB stars and differ by $\sim$ 2 mag 
in our investigated age ranges near [Fe/H] = $-$ 0.5. 
Therefore, well-calibrated $UBV$-band SBF magnitudes for 
old stellar systems can be a useful stellar population probe
as well as a distance indicator. In order for this system to be used for  
stellar population studies, more observations are 
clearly needed. It is especially important to obtain a $U$-band 
Galactic GC dataset to constrain the effects of post-RGB stars. 

Deriving relative ages from photometry alone is complicated by the fact 
that broadband colors suffer from age-metallicity degeneracy 
(e.g., G. Worthey 1994, ApJS, 95, 107). Our new models, 
which come from a consistent code, provide the 
tools to directly estimate ages of extragalactic GCSs using a 
variety of theoretical spectrophotometric quantities from far-UV to 
near-IR. Their use would clearly benefit this field 
of study. Well-calibrated separate metallicity information from 
spectroscopy, accompanied by precise spectrophotometry, is neccesary 
to extract the accurate age information for extragalactic GCSs. 
When the upcoming spectrophotometric observational data of GCSs 
in various morphological types of external galaxies from large ground-based 
telescopes and space-based UV telescopes are compared with 
our new models, we expect better determinations of the ages and metallicities. 
This accurate age information will be the most valuable 
for sorting out the several proposed present day galaxy formation scenarios 
(merger, accretion, in-situ, etc.). Furthermore, with the best available 
knowledge of star formation histories and chemical enrichment in 
galaxies, our models will become the building blocks of composite 
stellar population models. In the case of galaxy modeling, SBFs could 
be used as another constraint on stellar population studies in 
old stellar systems. 

\end{document}